\documentclass[aps,prb,twocolumn,showpacs]{revtex4-1}
\usepackage{graphicx}
\usepackage{amsmath}
\usepackage{amsfonts}
\usepackage{amssymb}
\usepackage{bm}
\usepackage{color}

\def\Im{\mathop{\rm Im}}
\def\Re{\mathop{\rm Re}}
\def\tr{\mathop{\rm tr}}
\def\Tf{\mathop{\rm \theta}}
\def\sign{\mathop{\rm sign}}

\def\be{\begin{equation}}
\def\ee{\end{equation}}

\newcommand{\corr}[1]{\left\langle #1 \right\rangle}

\newcommand{\bp}{\mathbf{p}}
\newcommand{\bq}{\mathbf{q}}
\newcommand{\bk}{\mathbf{k}}
\newcommand{\e}{\epsilon}
\newcommand{\ve}{\varepsilon}
\newcommand{\Om}{\Omega}
\newcommand{\om}{\omega}

\newcommand{\D}{\Delta}

\newcommand{\tom}{\tilde \omega}
\newcommand{\tD}{\tilde \Delta}

\begin{document}

\title{Isotope effect on the superfluid density in conventional and high-temperature superconductors}
\author{Maksym Serbyn and Patrick A. Lee}
\affiliation{Department of Physics, Massachusetts Institute of
Technology, Cambridge, Massachusetts 02139}
\date{\today}

\begin{abstract}
We investigate the isotope effect on the London penetration depth of a superconductor which measures $n_S/m^*$, the ratio of superfluid density to effective mass. We use a simplified model of electrons weakly coupled to a single phonon frequency  $\om_E$, but assume that the energy gap $\D$ does not have any isotope effect. Nevertheless we find an isotope effect for $n_S/m^*$ which is significant if $\D$ is sufficiently large that it becomes comparable to $\om_E$, a regime of interest to high $T_c$ cuprate superconductors  and possibly other families of unconventional superconductors with relatively high $T_c$. Our model is too simple to describe the cuprates and it gives the wrong sign of the isotope effect when compared with experiment, but it is a proof of principle that the isotope effect exists for $n_S/m^*$ in materials where the pairing gap and $T_c$ is not of phonon origin and has no isotope effect.
\end{abstract}


\maketitle
\section{Introduction}
While there is a general consensus that strong correlation governs the basic physics of high $T_c$ cuprates~\cite{LeeRevModPhys}, the role of electron-phonon interaction in determining $T_c$ is still under debate. The isotope effect is often viewed as an useful tool  which can provide information on this important issue. There is extensive isotope effect data on hole doped cuprates and the following picture has emerged. There exists significant isotope effect on the transition temperature for underdoped cuprates, but none for overdoped ones. On the other hand, for all doping substantial isotope effect on the London penetration depth $\lambda_{ab}$ has been observed~\cite{Khasanov03,Khasanov04}. Recall that $\lambda^{-2}_{ab}\propto n_S/m^*$ is a direct measure of the ratio between the superfluid density $n_S$ and the carrier mass $m^*$. The unusual isotope effect on $T_c$ can be understood qualitatively by the following picture~\cite{LeeRevModPhys}. For underdoped samples the transition temperature is controlled by the phase stiffness  $K_S=\hbar^2 n_S/4m^*$~\cite{Emery}. Hence, the isotope effect on $T_c$ may be simply inherited from the isotope effect on $n_S/m^*$. On the other hand, in overdoped samples $T_c$ is controlled by the pairing gap which has no isotope effect if phonons do not contribute significantly to its origin. Thus the isotope effect of $T_c$ can be qualitatively understood provided we accept the isotope effect on $n_S/m^*$. Then the puzzle is shifted to the origin of the isotope effect on $n_S/m^*$. Up to now there has been very little discussion in the literature on the isotope effect on $n_S/m^*$ in superconductors. In this paper we take the first step to address this important issue.

In~\cite{Khasanov03,Khasanov04} the authors suggested that the isotope effect on   $n_S/m^*$ is due to an isotope effect on the effective mass~$m^*$. However, the effective mass is given by $m^*/m=1+\lambda$ where~$\lambda$ is the dimensionless electron-phonon coupling.  Usually~$\lambda$ is not considered to have an isotope effect~\cite{McMillan}. This is because $\lambda=g^2N(0)\om_D$ and $g^2\approx\corr{I^2}/\om_D$, where $I$ is the coupling of the electron density to lattice deformation, which has no isotope dependence. This is supported by the direct experimental observation of Iwasawa \emph{et al.}~\cite{Iwasawa08} Using high-resolution laser ARPES measurements of electronic dispersion in Bi$_2$Sr$_2$CaCu$_2$O$_{8+\delta}$ (Bi2212), these authors study the influence of isotope oxygen substitution on the boson coupling ``kink'' in the electronic dispersion. The experiment clearly reveals an isotopic dependence of the kink energy, while it observes no change in the effective electron mass. On the other hand this experiment provides evidence for the coupling of phonon mode~\footnote{Authors of Ref.~[\onlinecite{Iwasawa08}] suggest that this is so-called breathing mode with $\Om\sim69$~meV coupled to the electrons with coupling parameter $\lambda\sim0.6$ that can be estimated from mass renormalization.} to electrons, which leads us to propose alternative explanation for the isotope effect on $n_S$.

Our model disregards any effects of strong correlation and considers a clean superconductor with $s$ or $d$-wave singlet pairing. We suppose that there exists one phonon mode weakly coupled to the electron system. However, the superconducting order parameter is not related to the phonon mode, but rather induced by some other mechanism. Therefore, we ignore the isotope effect on the gap $\D$ and $T_c$, even though these effects can easily be added to lowest order in $\lambda$. We assume the standard adiabatic approximation for electron-phonon coupling~\cite{[{We note that there is earlier works that ascribe the isotope effect to a change in hole density by appealing to the breakdown of the adiabatic approximation~[}]Kresin94}\phantom{\cite{*[{}][{]. We consider this scenario unlikely to be applicable to cuprates.}]Kresin98}} \!\!and neglect Coulomb interaction. The only quantity which depends on the isotope mass is $\om_E$.

\begin{figure}[t]
\hspace*{0.02\linewidth}%
\includegraphics[width=0.999\linewidth]{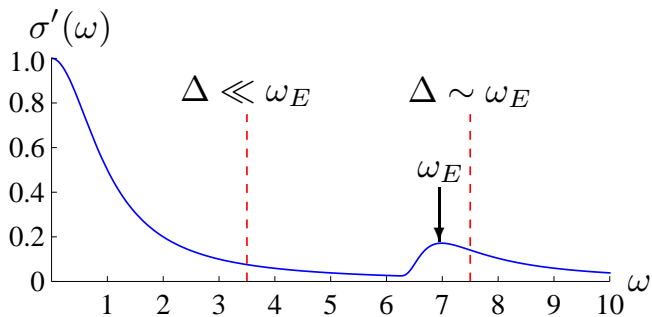}
\caption{\label{FigQual} Qualitative form of the conductivity for a normal system. With the onset of superconductivity, all spectral weight to the left of the corresponding red line goes to the condensate response.}
\end{figure}

First, we present simple qualitative arguments that explain the existence of the isotope effect on $n_S$ due to phonons using the sum rule which relates total spectral weight (SW) to the electron density and (bare) electron mass,
\begin{equation}
  \frac{ne^2}{m}
  =
  \frac{2}{\pi}\int_0^\infty d\om\,\sigma'(\om),
  \label{EqSumRule0}
\end{equation}
where $\sigma'(\om)$ is the dissipative part of the conductivity. The onset of superconductivity does not change the total SW, but is redistributed~\cite{Ferrell}. Some part of the SW now goes to the $\delta$-function response of the condensate, whose weight determines $n_S/m^*$, while the remaining goes to frequencies $\om\geq2\D$. This is visualized in the FIG.~\ref{FigQual}. It displays the dissipative part of conductivity for normal metal, $\sigma'(\om)$, that consists of a narrow Drude peak (since our sample is clean) along with a contribution from phonons at some characteristic frequency $\om_E$. Qualitatively with the onset of superconductivity some SW from frequencies $\om<\D$ goes to the $\delta$-peak and the SW from frequencies $\om>\D$ goes to the optical conductivity of superconductor. Thus, if we have $\D\ll\om_E$ (the case in weak coupling BCS theory), variation of $\om_E$ would have no effect on SW that goes to the condensate response. This may be the reason why this problem has not received much attention in the literature up to now. If $\D$ and $\om_E$ have the same order of magnitude, $n_S/m^*$ responds to the change of $\om_E$ and isotope effect for $n_S/m^*$ is present.

In the following sections we present calculations for the isotope effect for $n_S/m^*$ based on the sum rule. In Section~\ref{Sec2} we introduce our toy model, then calculate conductivity and superfluid density. In Section~\ref{Sec3} we discuss results and various approximations that were made. Finally, in the Appendix we present detailed calculations of the response kernel and conductivity.

\section{Calculation for $s$ and $d$-wave superconductors \label{Sec2}}

In this section we present our calculation of the isotope effect on the superfluid density $n_S$. We note, that both the experiment on the London penetration depth and the theory described in Eq.~(\ref{EqSumRule0}) give the combination $n_S/m^*$. From hereon we simply define $n_S$ as a product of $n_S/m^*$ and $m^*$ which is taken to be constant. We calculate the dissipative part of the conductivity due to phonons. After this, having the expression for the conductivity via the sum rule, we get access to the superfluid density.

The theoretical model is a superconductor with electron-phonon interaction, described by the standard  Fr\"olich Hamiltonian. The superconducting order parameter (singlet pairing with $s$ or $d$ wave gap symmetry) is assumed to be \emph{not related} to phonon mode under consideration. One may think about this pairing as induced by other phonon modes, strong correlation or some different mechanism. The resulting Hamiltonian can be written~as:
\be
  H
  =
  H_\text{BCS}+H_\text{ph}+H_\text{e-ph}.
\ee
$H_\text{BCS}$ is the BCS Hamiltonian, with the pairing in Cooper channel treated within mean-field theory (Coulomb interaction is neglected):
\begin{multline}
  H_\text{BCS}
  =
  \sum_\mathbf{p}\xi_\bp \Psi^+_\mathbf{p}\tau_3 \Psi_\mathbf{p}
  \\
  -
  \frac12\sum_{\mathbf{p}}
  \Psi^+_\bp (\D_\bp\tau_+ + \D^*_\bp \tau_-)\Psi_\bp,
\end{multline}
where $\xi_\bp$ is the Bloch energy measured relative to the Fermi energy (we consider only one electron band that is coupled to one phonon mode), $\tau^3$ is the Pauli matrix. Electron field operators here are written as a vector in the Nambu space
\be
  \Psi_\mathbf{p}
  =
  \left(
  \begin{matrix}
  c_{\mathbf{p}\uparrow}\\
  c^+_{-\mathbf{p}\downarrow}
  \end{matrix}
  \right),
\ee
with $c_{\bp,\uparrow}$, $c^+_{\bp,\uparrow}$ being standard creation and annihilation operators for up (down) spin electrons.
The Hamiltonian that describes the phonon mode and electron-phonon interaction is written as
\begin{multline}
  H_\text{ph}+H_\text{e-ph}
  =
  \sum_{\mathbf{k}} \om(\mathbf{k}) b^+_\mathbf{k}b_\mathbf{k}
  \\
  +
  \sum_{\mathbf{p},\mathbf{p}'} g(\bp,\bp')
  (b_\mathbf{\bp'-\bp}+b^+_\mathbf{\bp-\bp'})
  \Psi^+_{\bp'}\tau_3\Psi_{\bp}.
  \label{EqHFrolich}
\end{multline}
Here $\om(\mathbf{k})$ is  the (bare) phonon dispersion and  $g(\bp,\bp')$ is the  electron-phonon coupling. In what follows we use the Einstein approximation for the dispersion of phonons
\be
  \om(\bk)=\text{const}=\om_E,
\ee
and assume that our characteristic frequency $\om_E$ is on order of the superconducting gap, i.e. $\om_E\sim\D$ (note that for BCS superconductivity $\om_E\gg\D$). Moreover, we limit ourselves to leading order corrections in the electron-phonon coupling that is assumed to be weak. Thus we can neglect not only renormalization of the electron-phonon coupling~(that is small in the parameter $\om_E/E_F$~\cite{ Migdal}), but renormalization of the phonon dispersion as well.

\subsection{Green's function and self-energy\label{Sec2A}}
\begin{figure}[t]
\hspace*{0.02\linewidth}%
\includegraphics[width=0.999\linewidth]{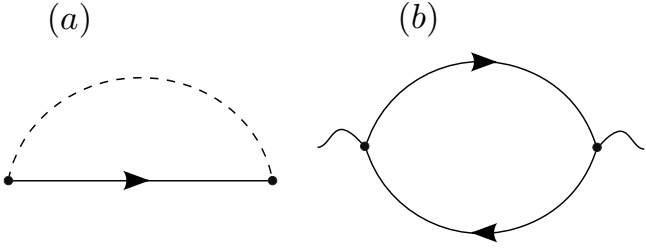}
\caption{\label{FigDiag} Diagrams corresponding to the self-energy~(a) corrections and conductivity~(b).}
\end{figure}
Before we start our calculations of conductivity we need to consider the electron Green's function. Following standard procedure, we define the Green's function of electrons $G_\bp(\tau)$~(which is a matrix in Nambu space) and phonons $D_\bq(\tau)$ in imaginary time as
\begin{gather}
  G_\bp(\tau)=-\corr{T_\tau(\Psi_\bp(\tau)\Psi_\bp^+(0))},
  \\
  D_\bq(\tau)=-\corr{T_\tau(b^+_\bp(\tau)b_\bp(0))}.
\end{gather}
Fourier transform of bare Green's functions (which are denoted by index $(0)$) can be written in the form~\cite{AGD}:
\begin{gather}
  (G^0_\bq(i\om_m))^{-1}
  =
  i\om_m - \xi_\bq \tau_3 - \Delta \tau_1,
  \label{EqGSC0Def}
  \\
  D^0_\bq(i\om_\nu)
  =
  \frac{1}{i\om_\nu-\om(\bq)}
  -
  \frac{1}{i\om_\nu+\om(\bq)},
\end{gather}
where $i\om_m=(2n+1)\pi T$ and $i\om_\nu=2\nu \pi T$
are the Matsubara frequencies for fermions and bosons.

In order to find the Green's function of electrons in the presence of the electron-phonon interaction we note that the superconducting order parameter is not related to the phonon mode under consideration. Moreover, the electron-phonon coupling is small and we limit ourselves to the leading order correction. Therefore, we can use perturbation theory to get access to the electron Green's function and there is no need in doing the full Eliashberg strong coupling theory~\cite{Eliashberg-60,*Eliashberg-61}.  Renormalization of the Green's function is given by the self-energy $\Sigma_\text{ph}$ due to electron-phonon interaction:
\begin{multline}
  (G^R_\bp(\om))^{-1}
  =
  (G^{0R}_\bp(\om))^{-1}-\Sigma_\text{ph}^R(\bp,\om)
  \\
  =
  {\tilde\om}^R(\om) - \xi_\bp \tau_3 - \tilde\Delta^R(\om) \tau_1,
  \label{EqGSCRenDef}
\end{multline}
where $\tilde\om (\om)$ and $\tilde \D (\om)$ are the renormalized~$\omega$ and gap~$\D$.

It suffices to calculate the self-energy to the \emph{leading order} in electron-phonon coupling. The expression for the self-energy in the Matsubara diagram technique can be read from the diagram depicted in FIG.~\ref{FigDiag}~(a):
\begin{multline}
  \Sigma_\text{ph}(\bp,i\om_n)
  =
  -T\sum_m
  \int(dp')
  |g(\bp,\bp')|^2
  \\
   D^0_{\bp-\bp'}(i\om_n-i\om_m)\tau^3 G^0_{\bp'}(i\om_m)\tau^3.
   \label{EqSigma1}
\end{multline}
Here the Pauli matrix $\tau_3$ (note that the self-energy is a matrix in the Nambu space) comes from the interaction vertex. Since we work to leading order of perturbation theory, we use the bare electron and phonon Green's functions. We proceed with analytical continuation~(the detailed procedure is described in \cite{Eliashberg-61,Scalapino}). Sums over positive and negative $m$ are represented as contour integrals. Afterwards, employing the spectral representation for the electron and phonon Green's functions~(see book~\cite{Scalapino}), one has:
\begin{multline}
  \Sigma^{R,A}_\text{ph}(\bp,\om)
  =
  -\int_{-\infty}^\infty (dx) \int_{0}^\infty d\Om\, \alpha^2(\Om)F(\Om)
  \\
  \int d\xi_{\bp'}
  \Im[\tau^3 G^{0R}_{\bp'}(x)\tau^3]
  \\
  \times
  \left[
  \frac{\tanh\frac{x}{2T}-\coth\frac{\Om}{2T}}{x-\om-\Om \mp i\delta}
  -
  \frac{\tanh\frac{x}{2T}+\coth\frac{\Om}{2T}}{x-\om+\Om \mp i\delta}
  \right],
  \label{EqSigma2}
\end{multline}
where $\alpha^2(\Om) F(\Om)$ is the spectral function of electron-phonon interaction (Eliashberg function) that consists of the product of effective electron-phonon coupling and phonon density of states. It is expressed through the electron-phonon interaction vertex as
\begin{multline}
  \alpha^2(\Om)F(\Om)
  =
  \frac{\int_{FS} d^2p
  \int_{FS} \frac{d^2p'}{(2\pi)^3v'_F} g(\bp,\bp')\delta(\Om-\om_{\bp-\bp'})}{\int_{FS}d^2p},
  \label{EqalphaF}
\end{multline}
where $v_F'$ is the Fermi velocity and integrations are over the Fermi surface. Finally, we integrate over the loop momentum~\footnote{Account for $\xi_\bp$ in the numerator of Green's function in the leading order gives the correction to the chemical potential that is small by parameter $\D^2/\e_F^2$, see Ref.~[\onlinecite{Scalapino}].}:
\begin{multline}
  \int d\xi_{\bp'}\, \Im[\tau^3 G^{0R}_{\bp'}(x)\tau^3]
  =
  -i\pi\sign x\frac{x+\tau^1 \D}{\sqrt{x^2-\D^2}},
\end{multline}
and put temperature $T=0$. Expanding $\Sigma^{R,A}_\text{ph}(\bp,\om)$ in components we obtain for renormalized $\om$ and $\D$, that were defined earlier in Eq.~(\ref{EqGSCRenDef}),
\begin{multline}
  {\tilde \om}^{R,A} (\om)
  =
  \om
  +
  \int_{-\infty}^\infty
  dx\,
  d\Om\,
  \alpha^2F(\Om)
  \Re\left[
  \frac{x \sign x}{\sqrt{x^2-\Delta^2}}
  \right]
  \\
  \left[
  \frac{\theta(-x)}{x-\om-\Om \mp i\delta}
  +
  \frac{\theta(x)}{z-\om+\Om \mp i\delta}
  \right]
  ,
  \label{EqZR}
\end{multline}
\begin{multline}
  {\tilde \Delta}^{R,A} (\om)
  =
  \Delta
  +
  \int_{-\infty}^\infty
  dx\,
  d\Om\,
  \alpha^2F(\Om)
  \Re\left[
  \frac{\Delta\sign x}{\sqrt{x^2-\Delta^2}}
  \right]
  \\
  \left[
  \frac{\theta(-x)}{x-\om-\Om \mp i\delta}
  +
  \frac{\theta(x)}{z-\om+\Om \mp i\delta}
  \right]
  .
  \label{EqDeltaR}
\end{multline}

It is convenient to introduce a dimensionless coupling constant $\lambda$ defined to be responsible for the electron mass renormalization due to phonons:
\be
  m^*
  =
  m(1+\lambda+O(\lambda)),
  \label{EqLambdaDef}
\ee
where $m$ is bare and $m^*$ is renormalized (measured in experiments, e.g. in ARPES~\cite{Iwasawa08}) mass. The dimensionless coupling is expressed through the Eliashberg function~(\ref{EqalphaF}) as:
\be
  \lambda
  =
  2
  \int \frac{d\om}{\om} \alpha^2(\om) F(\om).
  \label{EqlambdaDef}
\ee
We work in the Einstein approximation, where the dispersion of phonon mode does not depend on the momentum, $\om_\bp=\om_E$. Using this, one can immediately infer from Eq.~(\ref{EqalphaF}) that Eliashberg function $ \alpha^2(\Om) F(\Om)\propto\delta(\Om-\om_E)$. The constant of proportionality can be read from  Eq.~(\ref{EqlambdaDef}), giving us the expression for the Eliashberg function that will be used in the remainder of this paper:
\be
  \alpha^2(\Om) F(\Om)
  =
  \frac{\om_E\lambda}{2}
  \,
  \delta(\Om-\om_E).
\ee
In what follows we will make use of the imaginary part of the renormalized frequency and gap, that can be easily inferred from Eqs.~(\ref{EqZR})-(\ref{EqDeltaR}):
\begin{gather}
  {{\tilde \om}^R}{}''(\om)
  =
  \pi\om_E\lambda  \frac{\theta(|\om|-\om_E-\D)(|\om|-\om_E)}{2\sqrt{(|\om|-\om_E)^2-\D^2}},
  \label{EqtildeomIm}
  \\
  {\tilde \D}^{R}{}''(\om)
  =
  \pi\om_E\lambda \frac{\theta(|\om|-\om_E-\Delta)\Delta \sign\om}{2\sqrt{(|\om|-\om_D)^2-\D^2}},
  \label{EqtildeDIm}
\end{gather}
whereas, for advanced functions  we have
\begin{equation}
  {{\tilde \om}^A}{}''(\om)
  =
  -{{\tilde \om}^R}{}''(\om),
  \qquad
  {\tilde \D}^{A}{}''(\om)
  =
  -{\tilde \D}^{R}{}''(\om).
\end{equation}
Not surpisingly, the imaginary part of the self-energy is present only for $\om\geq\om_E+\D$, i.e. where the real exitation can be created. The square root singularity that is present at this threshold would be smeared for a more realistic phonon spectrum.
\subsection{Conductivity\label{Sec2B}}
Having the electron Green's functions at hand, we proceed to the calculations of the conductivity. Following the standard approach~\cite{AGD}, in order to calculate the conductivity we consider the response kernel $Q_{\alpha\beta}(\bk,\om)$ that relates current response to the vector potential:
\be
  j_\alpha(\bk,\om)
  =
  -\frac{ne^2}{m}Q_{\alpha\beta}(\bk,\om) A_\beta(\bk,\om).
  \label{EqQDefine}
\ee
The optical conductivity is expressed through $Q$ as:
\be
  \sigma_{\alpha\beta}(\bk,\om)
  =
  \frac{ne^2}{m}
  \frac{Q_{\alpha\beta}(\bk,\om)}{i\om}
  .
  \label{EqSigmaThroughQ}
\ee
The response kernel $Q_{\alpha\beta}(\bk,\om)$ can be calculated within the Kubo linear response method as a sum of the diamagnetic contribution and a current-current correlator~\cite{AGD}:
\be
  \frac{ne^2}{m}
  Q_{\alpha\beta}(\bk,\om)
  =
  \frac{ne^2}{m}\delta_{\alpha\beta}
   -
  {\cal P}^R_{\alpha\beta}(\bk,\om)
  ,
\ee
where ${\cal P}^R_{\alpha\beta}(\bk,\om)$ is analytic continuation to the real frequency of the (Fourier transformed) current-current correlator
\be
  {\cal P}_{\alpha\beta}(\mathbf{r}-\mathbf{r}',\tau-\tau')
  =
  \corr{T({\hat j}_{1\alpha}(\mathbf{r},\tau) {\hat j}_{1\beta}(\mathbf{r}',\tau'))}.
\ee
The current operator $\hat {\mathbf{j}}_{1}(\mathbf{r})$ is defined as a paramagnetic part of the full current operator $ \hat {\mathbf{j}}(\mathbf{r})$:
\begin{multline}
 \hat {\mathbf{j}}(\mathbf{r},\tau)
  =
  \frac{ie}{2m}(\nabla_{\mathbf{r}'}-\nabla_{\mathbf{r}})_{\mathbf{r}'\rightarrow \mathbf{r} }
  c^+(\mathbf{r}')c(\mathbf{r})
  \\
  -
  \frac{e^2}{m}\mathbf{A}(r)  c^+(\mathbf{r})c(\mathbf{r})
  \equiv
  \hat {\mathbf{j}}_{1}(\mathbf{r},\tau)
  -
  \frac{e^2}{m} \mathbf{A}(\mathbf{r})  c^+(\mathbf{r})c(\mathbf{r})
  .
  \label{EqjDef}
\end{multline}
Since we consider the superconductor in the London limit, we can put $\bk=0$. Denoting $Q_{xx}(0,\om)\equiv Q(\om)$ we proceed to the calculation of $Q(\om)$.

The current-current correlator ${\cal P}^R_{\alpha\beta}(k,\om)$ is given by the sum of all possible diagrams with 2 external vertices corresponding to current operators. If we neglect vertex corrections (see discussion in Section~\ref{Sec3}), in the leading order in the electron-phonon coupling we need to consider only the simplest diagram, FIG.~\ref{FigDiag}~(b). It is convenient to represent the diamagnetic contribution to $Q(\om)$ through Green's functions of a normal metal~\cite{AGD,Nam-67}. Combining the diamagnetic term with the current-current correlator given by the diagram in FIG.~\ref{FigDiag}~(b) we have:
\begin{multline}
  Q(i\om_n)
  =
  \frac12\tr\left[T\sum_{\om_m}\int d\xi_\bp\,
  (G_\bp(i\om_n+i\om_m) G_\bp(i\om_m)
  \right.
  \\
  -
  \left.
  G^{\D=0}_\bp(i\om_n+i\om_m) G^{\D=0}_\bp(i\om_m))
  \right].
  \label{EqQMatsubara}
\end{multline}
Starting from this expression we perform analytic continuation, and integrate over the loop momentum. Analytical continuation is done in the standard way and is presented in details in the literature~\cite{Nam-67,Scalapino} and in the Appendix. After lengthy but straightforward calculations we get the following expression for the imaginary part of the response kernel $Q(\om)$ at zero temperature:
\begin{multline}
  Q''(\om)
  =
  \frac12\Im
  \int_{\D-\om}^{-\D} dz\,
  \left[
   -\frac{g^{AA}(z+\om,z)-1}{\ve^A(z+\om)+\ve^A(z)}
   \right.
   \\
   \left.
  +
    \frac{g^{RA}(z+\om,z)+1}{\ve^R(z+\om)-\ve^A(z)}
  \right].
  \label{EqQpp}
\end{multline}
Where $\ve^{\alpha}(\om)$ with $\alpha=R,A$ is defined as:
\be
  \ve^{\alpha}(z)=\sign z \sqrt{(\tom^{\alpha}(z))^2-(\tD^{\alpha}(z))^2}
  ,
  \label{EqveRADef}
\ee
and $g^{\alpha\beta}$ are the structure factors introduced in~\cite{Nam-67}:
\be
  g^{\alpha\beta}(z_1,z_2)
  =
  \frac{{\tilde \om}^\alpha(z_1) {\tilde\om}^\beta(z_2) +\tD^\alpha(z_1)\tD^\beta(z_2)}{\ve^\alpha(z_1)\ve^\beta(z_2)}.
\ee
Using the expressions for the imaginary part of $\tom^{R,A}$ and $\tD^{R,A}$,  Eqs.~(\ref{EqtildeomIm})-(\ref{EqtildeDIm}), we expand to  leading order in the corrections to the Green's function. After some calculations shown in the Appendix we arrive at the final answer, that can be expressed via complete elliptic integrals of the second kind, $E(x)$ and $K(x)$:
\begin{multline}
 \sigma'(\om)
  =
  \frac{ne^2}{m}
  \frac{\pi\lambda\om_E\om_-}{\om^3}
  \\
  \left[
   E\left(\frac{\om^2_-}{\om^2_+}\right)
   \right.
   \left.
   -
   \frac{\D(\om-\om_E)}{\om_+^2} K\left(\frac{\om^2_-}{\om^2_+}\right)
   \right]
  \label{EqSigma1Final}
  ,
\end{multline}
where we used shorthand notation
\begin{gather}
  \om_\pm
  =
  \om - \om_E \pm 2\D.
\end{gather}
In case when there is no superconductivity in our system, putting $\D=0$ from~(\ref{EqSigma1Final}) we reproduce the well-known result for the leading order correction to the conductivity of normal metal due to phonons~\cite{Holstein}:
\be
  \sigma'_\text{N,e-ph}(\om)
  =
   \frac{n e^2}{m}
   \lambda
   \frac{\pi\om_E (\om-\om_E)}{\om^3}\theta(\om-\om_E).
  \label{EqSigmaD=0}
\ee
$\sigma'(\om)$ has been computed earlier by Allen~\cite{Allen-71}. However the second term in Eq.~(\ref{EqSigma1Final}) containing the elliptic function $K$ is missing in Allen's formula. This difference is negligible when $\D\ll\om_E$ but gives a small noticeable correction when $\D\gtrsim\om_E$.
\begin{figure}[t]
\hspace*{0.02\linewidth}%
\includegraphics[width=0.999\linewidth]{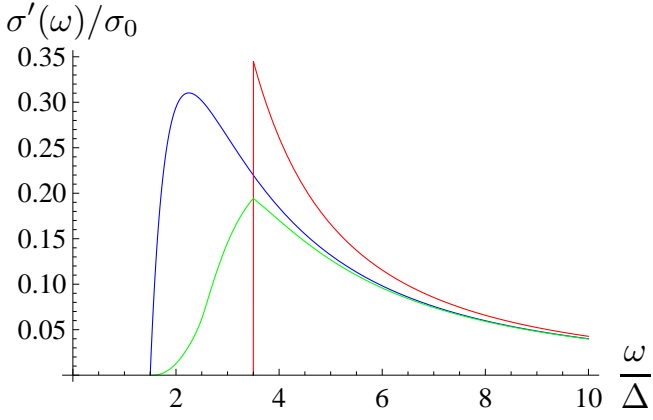}
\caption{\label{FigSigma} Correction to the dissipative part of conductivity $\sigma'(\omega)$ due to phonons for normal metal (blue) and superconductor with $s$ or $d$ wave symmetry (red and green curves correspondingly) for $\om_E=1.5\D$. Sigma normalized to $\sigma_0=\frac{ne^2}{m}\lambda$, Drude peak and $\delta$-function response of condensate are not shown.}
\end{figure}
\subsection{d-wave symmetry\label{Sec2C}}
The previous calculations can be easily generalized for the $d$-wave symmetry of pairing. We use the simplest possible model with the cylindrical Fermi surface with an axis parallel to the $c$-axis of a crystal. The order parameter is supposed to have $d_{x^2-y^2}$ structure. It can be written as
\be
  \D(\bk)
  =
  \D f
  \quad
  \text{with}
  \quad
  f=\cos 2\theta,
\ee
where $\theta$ is the angle between momentum (which lies in the $a-b$ plane) and $a$ axis.

First we recalculate our corrections to the electron Green's function. In contrary to the $s$-wave case, the gap $\Delta(\bk)$ is dependent on the direction of momentum. Therefore the previous answer for the $s$-wave case needs to be averaged over the Fermi surface. Averaging is defined as
\be
  \corr{X}_{FS}
  =
  \int_0^{2\pi} (d\theta)\, X(|\cos2\theta|)
  \label{EqDefAveragingDwave}
  .
\ee
Gap renormalization ${\tilde \D}^R(\om)-\D = 0$ vanishes under averaging over angle, while the imaginary part of the renormalization of $\om$ reads:
\begin{multline}
  {\tilde \om}^{R}{}'' (\om)
  =
  \frac{\pi\om_E\lambda}{2} \Tf(|\om|-\om_E)
  \corr{\frac{|\om|-\om_E}{\sqrt{(|\om|-\om_E)^2-\D^2f^2}}}_{FS}
  \\
  =
  \frac{\pi\om_E\lambda}{2} \Tf(|\om|-\om_E)\, \kappa\left(\frac{|\om|-\om_E}{\D}\right)
\end{multline}
where $\kappa(x)$ is defined as
\begin{equation}
  \kappa(x)
  =
  \frac2\pi
  \begin{cases}
  x K(x^2) \ \ \text{for} \ \ x\leq 1 \\
  K(x^{-2}) \ \ \text{for} \ \ x> 1
 \end{cases}
 .
\end{equation}

We repeat calculations of conductivity with $\tD(\om)=\D$ and the new expression for ${\tilde \omega}^{R}{}''(\om)$. Correction to the conductivity to leading order in the electron-phonon coupling is
\begin{multline}
 \sigma'(\om)
   =
 \frac{ne^2}{m}
  \frac{2}{\om^3}
 \corr{ \int_{\D|f|-\om}^{-\D|f|} dz\,
  \frac{z {\tilde \omega}^{R}{}''(z+\om)}{\sqrt{z^2-\D^2}}
  }_{FS}.
\end{multline}
Interchanging averaging over FS and integration over frequency we have two different expressions for
$\om_E\geq\D$ and $\om_E<\D$. When $\om_E\geq\D$ we obtain:
\begin{multline}
  \sigma'(\om)
  =
  \frac{ne^2}{m}
  \frac{\pi\om_E\lambda}{\om^3}
  \Tf(\om-\om_E)
  \\
  \int_{\om_E-\om}^{0} dz\,
   \kappa\left(\frac{|z+\om|-\om_E}{\D}\right)
   \kappa\left(\frac{|z|}{\D}\right)
 .
\label{EqSigma1dwave}
\end{multline}
While for $\om_E<\D$ we have a more complicated expression involving the elliptic integral of the first kind  $F(\sin\phi;k)$:
\begin{multline}
  \sigma'(\om)
   =
  \frac{ne^2}{m}
  \frac{\pi\om_E\lambda}{\om^3}
  \theta(\om-\om_E)
  \\
  \left[
  \int_{\D-\om}^{0} dz\,
   \kappa\left(\frac{|z+\om|-\om_E}{\D}\right)
   \kappa\left(\frac{|z|}{\D}\right)
   \right.
   \\
   \left.
   +
   \frac{2}{\pi}\int_{\om_E-\om}^{\D-\om} dz\,
   \kappa\left(\frac{|z+\om|-\om_E}{\D}\right)
   F\left(\frac{z+\om}{\D};\frac{\D^2}{z^2}\right)
 \right]
 .
 \label{EqSigma1dwave2}
\end{multline}

\subsection{Isotope effect for superfluid density $n_S$ \label{Sec2D}}
\begin{figure}[t]
\hspace*{0.02\linewidth}%
\includegraphics[width=0.999\linewidth]{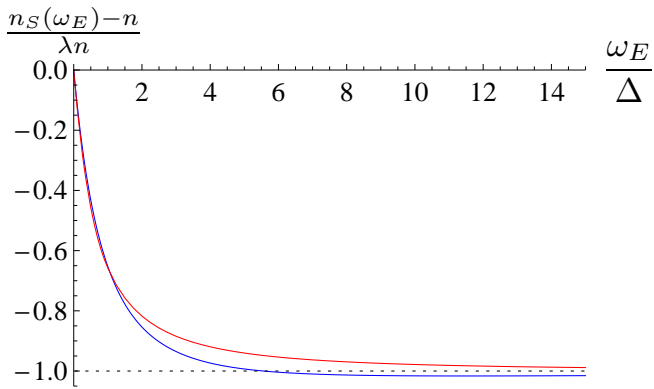}
\caption{Correction to superfluid density for superconductor with $s$-wave~(blue) or $d$-wave~(red) symmetry.
\label{FignS}}
\end{figure}
Using our explicit results for the conductivity, by means of the sum rule, we get access to the superfluid density. Initially the sum rule is formulated as
\be
  \frac{ne^2}{m}
  =
  \frac{2}{\pi}\int_0^\infty d\om\,\sigma'(\om),
  \label{EqSumRule}
\ee
where $n$ and $m$ are mass and concentration of electrons, while $\sigma'(\om)$ is the dissipative part of the conductivity.

Let us first apply the sum rule to our system when there is no superconductivity, i.e. $\D=0$. We assume that the mean free path is big, thus scattering time due to impurities  $\tau$ satisfies $\tau\om_E\gg 1$. Then, in addition to the very narrow Drude peak at $\om=0$, $\sigma_\text{D}(\om)$, we have the small correction due to phonons, $\sigma_\text{N,e-ph}'(\om)$, that starts at $\om_E$~(see Eq.~(\ref{EqSigmaD=0})). The calculation of the spectral weight of the Drude peak from the sum rule goes as follows:
\begin{multline}
   \frac{2}{\pi}\int_0^\infty d\om\,\sigma_\text{D}'(\om)
   =
    \frac{ne^2}{m}
    -
    \frac{2}{\pi}\int_0^\infty d\om\,\sigma_\text{N,e-ph}'(\om)
    =
    \\
   = \frac{ne^2}{m} - \lambda \frac{ne^2}{m}
   =
  \frac{ne^2}{m^*}(1 + O(\lambda^2)),
  \label{EqSumRuleN}
\end{multline}
where $m^*=m(1+\lambda)$, see Eq.~(\ref{EqLambdaDef}). We see that the SW of $\sigma_\text{N,e-ph}'(\om)$ does not depend on $\om_E$ and the sum rule reproduces the correct answer for the spectral weight of the Drude peak with the renormalized mass.

With the onset of the superconductivity the picture does not change drastically. The dissipative part of the conductivity consists of the $\delta$-function with the weight given by $n_Se^2/m^*$~(response of the condensate) and the small contribution due to electron-phonon interaction $\sigma'(\om)$~(calculated in Eq.~(\ref{EqSigma1Final}) and Eqs.~(\ref{EqSigma1dwave})-(\ref{EqSigma1dwave2}) for $s$ and $d$-wave pairing). The sum rule implies for $n_S$:
\be
  n_S
  =
  n-\frac{2m}{\pi e^2}
  \int^\infty_{2\D} d\om\, \sigma'(\om).
  \label{EqSumRulenS}
\ee
From Eqs.~(\ref{EqSigma1Final}) and (\ref{EqSigma1dwave}), the superfluid density $n_S$ can be easily computed. It depends on $\om_E$, thus giving nonzero isotope effect on superconducting density. Results of the numerical computations of $(n_S-n)/\lambda$ as a function of $\om_E/\D$ are shown in  FIG.~\ref{FignS}.

\begin{figure}[t]
\hspace*{0.02\linewidth}%
\includegraphics[width=0.999\linewidth]{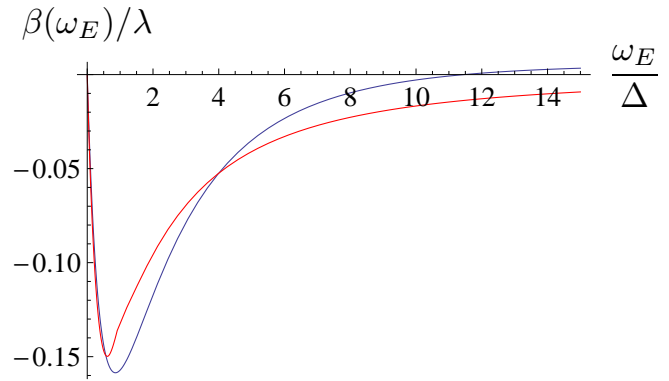}
\caption{\label{FigBeta} Isotope coefficient $\beta$ defined in main text as a function of $\om_E$ in our model.}
\end{figure}

Based on the dependence of $n_S$ on $\om_E$, an isotope coefficient  $\beta$  can be easily calculated. It is defined as~\cite{Khasanov03}
\be
  \beta=-\frac{d\log \lambda_{ab}^{-2}}{d\log M_\text{O}}
  =
  \frac12\frac{d\log \lambda_{ab}^{-2}}{d\log \om_E}
  \label{EqBetaDef},
\ee
where the usual relation between the phonon characteristic frequency and the atomic mass $M_\text{O}$~(in experiment this is usually the atomic mass of oxygen) was assumed:
\be
  \frac{\D \om_E}{\om_E}
  =
  -\frac12\frac{\D M_\text{O}}{M_\text{O}}.
\ee
Results for $\beta$ are shown in the FIG.~\ref{FigBeta}. Notably, for both $s$ and $d$-wave symmetry, the isotope coefficient has maximum modulus when $\om_E\sim\D$. While when $\om_E\gg\D$, $\beta$ vanishes in either case. Such behavior naturally follows from our qualitative arguments, presented in the Introduction. However note that $\beta$ approaches zero from positive and negative values for the $s$-wave and $d$-wave pairing respectively.

\section{Discussion \label{Sec3}}
\begin{figure*}[t]
\hspace*{0.02\linewidth}%
\includegraphics[width=0.999\linewidth]{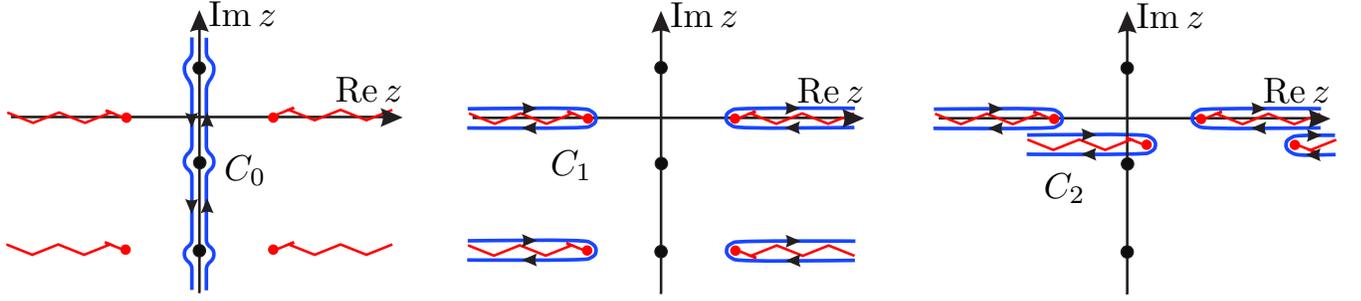}
\caption{\label{FigContour} Deformation of integration contour during analytical continuation. Branch cuts of Green's functions are shown in red and starts at $\om_\D$ and $\om_\D-i\om_m$. When we do analytical continuation from $\om_n \rightarrow -i\om$, all four branch cuts lie on the real axis.}
\end{figure*}
Although our toy model proves that isotope effect on the superfluid density can exist due to phonons, it gives an effect of the opposite sign to the experimentally measured. For example in~\cite{Khasanov03} for YPrBaCuO  $\beta=0.38$ for $x=0.3$ and  $\beta=0.71$ for $x=0.4$, while our model predicts the isotope coefficient of order of $\beta \sim -0.1$ for $\lambda\sim1$.

The opposite sign of isotope coefficient leads us to a discussion of the limitations of our model.  We treated the electron-phonon interaction as being weak which is not really the case. As it can be inferred from the experiment~\cite{Iwasawa08,Note1}, the coupling constant is of order one. However, the qualitative features of our result should not change. The second approximation  was the neglect of vertex corrections. The justification of this has been given in~\cite{Allen-71}. Since for normal metal ladder diagrams give minor effects, we expect this to be the case for superconductors as well.

The most serious limitation of our model with respect to high $T_c$ superconductors is the neglect of strong correlation effects. The minimum requirement would be to reproduce the fact that $n_S$ is proportional to the doping hole concentration $x$ due to the proximity to the Mott transition, and not to the electron density $1-x$ as in our toy model.  Furthermore, from the point of view of the sum rule argument, it is known that strong correlation gives rise to an incoherent background at finite frequencies, the so-called mid-infrared peak, even before phonons are taken into account. This incoherent background is missing in our toy model. What is needed is clearly a model which includes both Coulomb repulsion and coupling to phonons, such as the Hubbard-Holstein model. There has been considerable progress in this difficult problem, as given in a recent review~\cite{Castellani}. It may be possible to extend the present model to include correlation effect in the future.
Even though our result does not agree with the experiment, the important message of the present paper is a proof of principle that isotope effect in $n_S/m^*$ is possible even if the pairing is not due to phonons.

Finally the isotope effect on the superfluid density discussed in this paper should be applicable to conventional superconductors in the clean limit with minor extensions. To linear order in $\lambda$, it is straightforward to include coupling to a distribution of phonon modes. Furthermore, we can easily include the isotope effect on the energy gap $\D$ in our consideration. In weak coupling BCS theory, $\D$ is proportional to $\om_E$ and the ratio $\D/\om_E$  has no isotope dependence. Then our theory trivially predicts no isotope effect on $n_S/m^*$.
Including $\mu^*=\mu/\log(\epsilon_F/\om_E)$ in the $T_c$ formula will introduce some isotope dependence in $\D/\om_E$. However, in most superconductors $\om_E/\D\gg1$ and from FIG.~\ref{FigBeta} we see that the predicted effect is very small. Recently $s$-wave superconductors with relatively large $T_c$ and energy gap have been discovered. Examples are MgB$_2$ and doped fullerenes. While coupling to certain high frequency phonons may be responsible for the pairing, there exist in these materials lower frequency phonons with frequency $\om_0$ which may bring us to the regime of intermediate $\om_0/\D$. It will be interesting to search for the isotope effect on the penetration depth in these materials (provided they are in the clean limit). In addition, the new class of Fe-based superconductors have a rather large ratio of $\D/\om_E$ and should exhibit the isotope effect on the penetration depth according to our theory.

\appendix
\section{\label{SecApp1} Derivation of the response kernel and conductivity}

In this Appendix, following~\cite{AGD,Nam-67} we derive the response kernel $Q(\om)$ and dissipative part of the optical conductivity $\sigma'(\om)$ due to the electron-phonon interaction. Starting from the expression for $Q(i\om_n)$, Eq.~(\ref{EqQMatsubara}), we do analytic continuation to the real frequencies. Performing integration over $\xi_\bp$ and setting temperature to zero, we arrive to the final expression for the response kernel. Afterward, using corrections to the self-energy and expanding in the electron-phonon coupling we get the conductivity.

Matsubara sum in~(\ref{EqQMatsubara}) is represented as a contour integral in the standard way:
\begin{multline}
  \sum_{\om_m}
  G_\bp(\om_n+\om_m) G_\bp(\om_m)
  =
  \\
  \frac{1}{4\pi i}\int_{C_0}dz\,
  \tanh \frac{z}{2T} G_\bp(\om_n-iz) G_\bp(-iz),
  \label{EqMsum}
\end{multline}
where the contour $C_0$ surrounds poles of $\tanh \frac{z}{2T}$ that lie on the imaginary axis (see FIG.~\ref{FigContour}). Having in mind subsequent integration over $\xi_\bp$, it is convenient to represent the electron Green's function~(\ref{EqGSCRenDef}) as~\cite{Nam-67}:
\begin{multline}
  G_\bp(i\om_n)
  =
  \frac{1}{2\ve(i\om_n)}
  \left[
  \frac{\tom(i\om_n)+\tau^3\ve(i\om_n)+\tau^1\tD(i\om_n)}{\ve(i\om_n)-\xi_\bp}
  \right.
  \\
  \left.
  +
  \frac{\tom(i\om_n)-\tau^3\ve(i\om_n)+\tau^1\tD(i\om_n)}{\ve(i\om_n)+\xi_\bp}
  \right]
  ,
  \label{EqAppGRenRepresentation}
\end{multline}
where $\ve(\om)$ is defined as
\be
  \ve(\om)=\sqrt{\tom^2(\om)-\tD^2(\om)}
  .
  \label{EqveDef}
\ee
Originally $G_\bp(i\om_n)$ in Eq.~(\ref{EqAppGRenRepresentation}) is defined only at the discrete set of Matsubara frequencies on the imaginary axis. Analytic properties of $G_\bp(z)$ in the complex plane are related to the properties of function $\ve(z)$. We suppose that on the imaginary axis $\ve(z)$ is a well defined function. Let us denote a point where  $\ve(z)$ has an essential singularity as $\om_\D$. We argue that $\om_\D$ is real. Indeed, on the real axis the self-energy correction has a non-zero imaginary part only for frequencies above the threshold, $\om\geq\D+\om_E$. Therefore, $\om_\D$ is real and $\om_\D=\D+O(\lambda)$, where $O(\lambda)$ -- corrections of order of $\lambda$. Drawing branch cuts from points $\pm\om_\D$ to infinity (they are shown in red zigzags in FIG.~\ref{FigContour}) we make $\ve(z)$ a well defined function in the whole complex plane.

Retarded and advanced Green's functions $G_\bp^{R,A}(z)$ are given by the value of $G_\bp(i\om_n)$ at the upper (lower) side of the branch cut. They are obtained from Eq.~(\ref{EqAppGRenRepresentation}) by replacing functions $\tom(z)$, $\tD(z)$, and $\ve(z)$ with their analytic continuation to the upper (lower) side of cuts. These are denoted as $\tom^{R,A}(z)$, $\tD^{R,A}(z)$, and $\ve^{R,A}(z)$ correspondingly, where first two functions were calculated in the Section~\ref{Sec2A},
and last function is defined as
\be
  \ve^{\alpha}(z)=\sign z \sqrt{(\tom^{\alpha}(z))^2-(\tD^{\alpha}(z))^2}
  ,
  \label{EqAppveRADef}
\ee
where $\alpha=R, A$.
Now we deform the contour from $C_0$ to $C_1$ (see FIG.~\ref{FigContour}) and analytically continue to the real external frequency $\om_n \rightarrow -i\om$. As a result, the Matsubara sum~(\ref{EqMsum}) is written as
\begin{widetext}
\begin{multline}
  \sum_{\om_m}
  G_\bp(\om_n+\om_m) G_\bp(\om_m)
  =
  \frac{1}{4\pi i}\int^\infty_{\om_\D}dz\,
  \tanh \frac{z}{2T}
  \Big\{
  (G_\bp^R(z)-G^A_\bp(z))(G^R_\bp(z+\om)+G^A_\bp(z-\om))+
  \\
  (G^R_\bp(-z)-G^A_\bp(-z))(G^R_\bp(-z+\om)+G^A_\bp(-z-\om))
  \Big\}.
  \label{EqMsumInt}
\end{multline}
The trace of the product of two Green's functions $ G^\alpha(z_1) G^\beta(z_2)$ with $\alpha,\beta=R,A$ can be written as:
\begin{multline}
  2\tr G^\alpha(z_1) G^\beta(z_2)
  =
  \frac{g^{\alpha\beta}(z_1,z_2)+1}{[\ve^\alpha(z_1)-\xi_\bp][\ve^\beta(z_2)-\xi_\bp]}+\frac{g^{\alpha\beta}(z_1,z_2)+1}{[\ve^\alpha(z_1)+\xi_\bp][\ve^\beta(z_2)+\xi_\bp]}
  +
  \\
  \frac{g^{\alpha\beta}(z_1,z_2)-1}{[\ve^\alpha(z_1)-\xi_\bp][\ve^\beta(z_2)+\xi_\bp]}+\frac{g^{\alpha\beta}(z_1,z_2)-1}{[\ve^\alpha(z_1)+\xi_\bp][\ve^\beta(z_2)-\xi_\bp]}
  ,
  \label{EqTrGG}
\end{multline}
\end{widetext}
where $g^{\alpha\beta}$ are the structure factors introduced in~\cite{Nam-67} as
\be
 g^{\alpha\beta}(z_1,z_2)
  =
  \frac{{\tilde \om}^\alpha(z_1) {\tilde\om}^\beta(z_2) +\tD^\alpha(z_1)\tD^\beta(z_2)}{\ve^\alpha(z_1)\ve^\beta(z_2)}.
\ee
While integrating over $\xi_\bp$, we close the integration contour in the upper half plane and use representation~(\ref{EqTrGG}) along with the property $\Im\ve^{R}(z)> 0$, $\Im\ve^{A}(z)< 0$. For example, integration of the product of two retarded Green's functions from Eq.~(\ref{EqMsumInt}) yields
\begin{equation}
  \int d\xi_\bp \tr\left[G_\bp^R(z)G_\bp^R(z+\om)\right]
  =
  -2\pi i \frac{g^{RR}(z,z+\om)-1}{\ve^R(z)+\ve^R(z+\om)}.
\end{equation}
Gathering contributions from all terms and using the symmetry
$ \ve^R(-z) = -\ve^A(z)$, we have:
\begin{multline}
  Q(\om)
  =
  \frac12\int_{\om_\D}^\infty dz\,
  \left[\tanh \frac{z+\om}{2T} B -\tanh \frac{z}{2T} A \right]
  \\
  +
  \frac12\int_{\om_\D-\om}^{\om_\D} dz\,
  \tanh \frac{z+\om}{2T} B,
  \label{EqAppQ}
\end{multline}
where
\begin{gather}
  A
  =
  \frac{g^{RR}(z+\om,z)-1}{\ve^R(z+\om)+\ve^R(z)}
  +
    \frac{g^{RA}(z+\om,z)+1}{\ve^R(z+\om)-\ve^A(z)},
\\
  B
  =
  -\frac{g^{AA}(z+\om,z)-1}{\ve^A(z+\om)+\ve^A(z)}
  +
    \frac{g^{RA}(z+\om,z)+1}{\ve^R(z+\om)-\ve^A(z)}.
    \label{EqAppB}
\end{gather}
We restrict ourselves to zero temperature. $B-A$ is real, thus it does not contribute to the imaginary part of the response kernel, and consequently to $\sigma'(\om)$,
\begin{equation}
  \sigma'(\om)
  =
  \frac{ne^2}{m}
  \frac{Q''(\om)}{\om}
  .
  \label{EqAppSigmaQ}
\end{equation}
The region of integration from $-\om_\D$ to $\om_\D$ in the second integral in Eq.~(\ref{EqAppQ}) also gives a real contribution and can be omitted. Finally, neglecting the difference between $\om_\D$ and $\D$  that gives higher order corrections in $\lambda$, finally we have for $Q''(\om)$:
\begin{equation}
  Q''(\om)
  =
  \frac12
  \int_{\D-\om}^{-\D} dz\,
  \Im B
  \label{EqAppQpp}
\end{equation}
which coincide with Eq.~(\ref{EqQpp}) used in Section~\ref{Sec2B}.

In order to calculate the conductivity we use the imaginary part of $\tom^{R,A}(z)$ and $\tD^{R,A}(z)$, Eqs.~(\ref{EqZR})-(\ref{EqDeltaR}). We work to leading order in the coupling constant, therefore we expand $B$ in Eq.~(\ref{EqAppQpp}) and extract its imaginary part.

An imaginary contribution to $B$ comes only from the imaginary corrections to the self-energy, $\tom^{R}{}''(z)$ and $\tD^{R}{}''(z)$, Eqs.~(\ref{EqtildeomIm})-(\ref{EqtildeDIm}). They are proportional to the coupling constant, thus we expand in them.  First, we expand $\ve^{R,A}(z)$:
\begin{equation}
  \ve^{R,A}(z)
  =
  \e(z) \sign z   \pm  i\Gamma^R(z),
\end{equation}
with short-hand notation
\be
  \e(z)
  =
  \sqrt{z^2-\D^2},
\ee
\be
  \Gamma^R(z)
  =
  \pi\om_E\lambda \frac{\theta(|z|>\D+\om_D) |z|(|z|-\om_D)-\D^2)}{2\sqrt{z^2-\D^2}\sqrt{(|z|-\om_D)^2-\D^2}}.
\ee
Using the expansion of $\ve^{R,A}(z)$ we expand the structure factors and numerators in~(\ref{EqAppB}). After some calculation we get
\begin{widetext}
\begin{multline}
  \frac12\Im B
  =
  -\frac{\D }{\om\e (z) \e(z+\om)}
  \left[\frac{\D \tom^R{}''(z) + z \tD^R{}''(z)}{\e(z)} + \frac{\D \tom^R{}''(z+\om) + (z+\om) \tD^R{}''(z+\om)}{\e(z+\om)}\right]
  \\
  +
   \frac{z(z+\om)-\D^2}{\om^2\e(z)\e(z+\om)}
  \left[   \Gamma^R(z)+\Gamma^{R}(z+\om)   \right].
\end{multline}
To simplify this further we use the explicit form of $\tom^{R''}(z)$, $\tD^{R''}(z)$, and $\Gamma^R(z)$.
Inserting $\Im B$ into  Eqs.~(\ref{EqAppSigmaQ})-(\ref{EqAppQpp}) and changing integration variables yields
\begin{equation}
 \sigma'(\om)
  =
  \frac{ne^2}{m}
  \frac{\pi\om_E\lambda}{\om^3}
  \\
  \int_{-\gamma+\D}^{\gamma-\D} dt\,
  \frac{(-t+\gamma)(t-\gamma)+\D^2}{\sqrt{(t+\gamma)^2-\D^2}\sqrt{(t-\gamma)^2-\D^2}}
  ,
\end{equation}

where $\gamma=(\om-\om_D)/2$. This integral can be expressed through elliptic functions, resulting in Eq.~(\ref{EqSigma1Final}).
\end{widetext}
\bibliography{SerbynLeeIEhTc}
\end{document}